\documentclass[12pt]{article}
\usepackage[latin9]{inputenc}
\usepackage{amsbsy}
\usepackage{amssymb}
\usepackage{graphicx}
\usepackage{esint}
\PassOptionsToPackage{normalem}{ulem}
\usepackage{ulem}

\makeatletter

\providecommand{\tabularnewline}{\\}

\@ifundefined{date}{}{\date{}}

\usepackage{scicite}

\usepackage{times}

\newcounter{lastnote}

\newcounter{twocoltable}[table]

\newcounter{twocolfigure}[figure]

\renewenvironment{abstract}{\begin{quote} \bf}{\end{quote}}
\usepackage{jour-abbrev}
\usepackage{color}
\definecolor{shadecolor}{cmyk}{0.2,0.0,0.5,0.0}

\usepackage{multibbl}
\newbibliography{M} 
\newbibliography{S} 
\usepackage{fancyheadings}

\makeatother

\begin{document}
\pagestyle{fancyplain}
\chead{T. Alexander, P. Natarajan, Science, 7 August 2014 (10.1126/science.1251053).}

\bibliographystyle{M}{Science} 
\bibliographystyle{S}{Science}

\global\long\def\Mo{M_{\odot}}
\global\long\def\Ro{R_{\odot}}
\global\long\def\Mbh{M_{\bullet}}
\global\long\def\Ms{M_{\star}}
\global\long\def\Rs{R_{\star}}
\global\long\def\Ns{N_{\star}}
\global\long\def\ra{r_{a}}
\global\long\def\sigs{\sigma_{\star}}
\global\long\def\sbh{\sigma_{\bullet}}
\global\long\def\dbh{\Delta_{\bullet}}
\global\long\def\ddbh{\dot{\Delta}_{\bullet}}
\global\long\def\vbh{v_{\bullet}}
\global\long\def\rbh{r_{\bullet}}
\global\long\def\vcirc{v_{\phi}}
\global\long\def\vrel{v_{\mathrm{rel}}}
\global\long\def\Medd{\dot{M}_{E}}
\global\long\def\Mdot{\dot{M}}
\global\long\def\rtr{r_{\gamma}}
\global\long\def\dinf{\rho_{\infty}}
\global\long\def\cinf{c_{\infty}}
\global\long\def\tinf{t_{\infty}}
\global\long\def\taui{\tau_{a}^{0}}
\global\long\def\taua{\tau_{a}}
\global\long\def\Mtau{\dot{M}_{\tau}}
\global\long\def\tk{t_{\kappa}}
\global\long\def\jiso{j_{\mathrm{ISO}}}
\global\long\def\epsr{\epsilon_{\rho}}
\global\long\def\epsu{\epsilon_{u}}

\title{Rapid growth of seed black holes in the early universe\\
by supra-exponential accretion}

\author{Tal Alexander,$^{1\star}$ Priyamvada Natarajan$^{2}$\\
{\normalsize{}$^{1}$Department of Particle Physics \& Astrophysics,
Weizmann Institute of Science, Rehovot 76100, Israel}\\
{\normalsize{}$^{2}$Department of Astronomy, Yale University, 260
Whitney Avenue, New Haven, CT 06511, USA}\\
{\normalsize{}$^{\star}$Corresponding author. E-mail: tal.alexander@weizmann.ac.il}}
\maketitle
\begin{abstract}
Mass accretion by black holes (BHs) is typically capped at the Eddington
rate, when radiation's push balances gravity's pull. However, even
exponential growth at the Eddington-limited $e$-folding time $t_{E}\sim\mathrm{few\times}0.01$
billion years, is too slow to grow stellar-mass BH seeds into the
supermassive luminous quasars that are observed when the universe
is $1$ billion years old. We propose a dynamical mechanism that can
trigger supra-exponential accretion in the early universe, when a
BH seed is bound in a star cluster fed by the ubiquitous dense cold
gas flows. The high gas opacity traps the accretion radiation, while
the low-mass BH's random motions suppress the formation of a slowly-draining
accretion disk. Supra-exponential growth can thus explain the puzzling
emergence of supermassive BHs that power luminous quasars so soon
after the Big Bang.
\end{abstract}
Optically bright quasars powered by accretion onto black holes (BHs)
are now detected at redshifts as high as $z\sim7$, when the Universe
was 6\% of its current age ($<1$ billion years) \cite{M}{mor+11}.
Their luminosities imply supermassive BHs (SMBHs) with the mass of
the BH ($\Mbh$) $\gtrsim10^{9}$ solar masses ($\Mo$) \cite{M}{fan+06}.
The main obstacles to assembling such SMBHs so rapidly are the low
masses of the hypothesized initial seed BHs, born of first-generation
(Pop~III) stars, coupled with the maximal growth rate for radiatively
efficient accretion, the Eddington limit \cite{M}{jeo+12,mil+09,par+12}.
Proposed ways to circumvent these limitations invoke super-Eddington
accretion for brief periods of time \cite{M}{vol+05b}; the \emph{ab-initio}
formation of more massive BH seeds \cite{M}{hop+07b,vol10,nat11,hai13}
from the direct collapse of self-gravitating pre-galactic gas disks
at high redshifts \cite{M}{bro+03,lod+06,beg+06,lod+07}; and the
formation of a very massive star from runaway stellar mergers in a
dense cluster \cite{M}{dev+09,dav+11}. Discriminating between
these scenarios is challenging, because seed formation redshifts ($z>10$)
are observationally inaccessible. Current data require finely tuned,
continuous early BH growth and massive initial BH seeds \cite{M}{wil+03,fer+13,joh+13,tre+13}.
Recent results from high-resolution simulations of early star formation
at $z\sim15$ to $18$ exacerbate the problem by indicating that efficient
fragmentation and turbulence \cite{M}{alv+09,gre+11,tur+12,reg+14,saf+14}
lead to the efficient formation of stellar clusters embedded in the
flow, which prevents the formation of massive seeds ($\gg10\,\Mo$)
by limiting the mass of their potential Pop~III progenitor stars.
On the other hand, theoretical and numerical results on larger scales
suggest that ubiquitous dense cold gas flows \cite{M}{dek+09}
stream in along filaments and feed proto-galactic cores \cite{M}{dub+12,bou+11}.
Adaptive mesh refinement simulations track the fate of these sites
(collapsed $10^{7}\,\Mo$ dark matter halos) from $1\,\mathrm{Mpc}$
scale at $z\sim21$ with resolutions as low as $\sim2\times10^{-10}$
pc in the central regions. These simulations find isothermal density
cusps that reach extreme central densities, with an average density
of $\dinf\gtrsim10^{-16}\,\mathrm{g\, cm^{-3}}$ ($\gtrsim10^{8}\,\mathrm{cm^{-3}}$
for pure H) on $0.1$ pc scales \cite{M}{wis+08}. They also reveal
a marginally unstable central gas reservoir of $\mathrm{few\times}10^{5}\,\Mo$
in the inner few parsecs \cite{M}{footnote1} , where the dynamical
timescale is $\sim10^{6}$ years. Although these simulations are somewhat
idealized, we adopt the properties of this high density environment
as the initial conditions for the model presented here.

\begin{figure}[!p]
\centering{}\includegraphics[width=1\columnwidth]{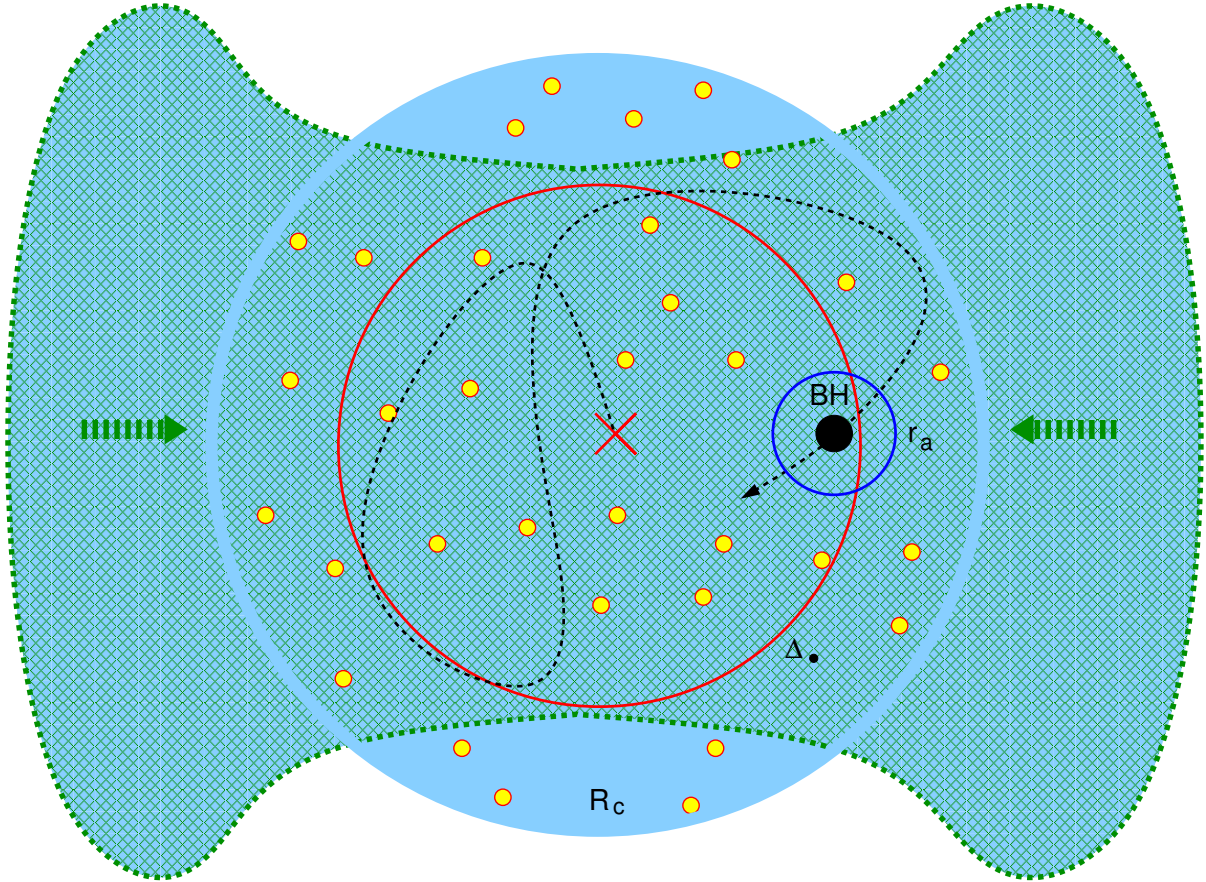}
\protect\caption{\label{f:M.schematic}A schematic depiction of accretion by a low-mass
BH in a dense gas-rich cluster. Dense cold gas (green) flows to the
center (red cross) of a stellar cluster (light blue region) of total
mass $M_{c}=\protect\Ns\protect\Ms+M_{g}$ and radius $R_{c}$, which
contains $N_{\star}$ stars (yellow circles) of mass $\protect\Ms$
each with velocity dispersion $\protect\sigs$, and gas of mass $M_{g}$.
The gas is nearly pressure-supported and close to the virial temperature.
A stellar BH (black circle) of mass $\protect\Ms<\protect\Mbh\ll M_{c}$,
which is accreting from its capture radius $r_{a}$ (dark blue circle),
is initially in fluctuation-dissipation equilibrium with the stars
and is scattered by them (black dashed line) with velocity dispersion
$\protect\sbh\sim\sqrt{\protect\Ms/\protect\Mbh}\protect\sigs$ over
a distance scale $\protect\dbh\sim\sqrt{\protect\Ms/\protect\Mbh}R_{c}$
(red circle). }
\end{figure}

We consider a scenario in which a low-mass Pop~III remnant BH remains
embedded in a nuclear star cluster fed by dense cold gas flows \cite{M}{dek+09}
(Figure~\ref{f:M.schematic} and Table~\ref{t:M.model}). The stars
and gas are virialized in the cluster potential, and the BH is initially
a test particle in equipartition with the stars. Gas within the accretion
(capture) radius of the BH, $r_{a}=[2c^{2}/(\cinf^{2}+\vbh^{2})]r_{g}$,
is dynamically bound to it, where $r_{g}=G\Mbh/c^{2}$ is the gravitational
radius of the BH; $\cinf$ is the gas sound speed in the cold flow
far from the BH, which is a measure of the depth of the cluster's
gravitational potential; and $\vbh$ is the BH velocity relative to
the gas. Gas bound to the BH inside $r_{a}$ is not necessarily accreted
by it. Prompt accretion requires gas to flow from $r_{a}$ into the
BH on a plunging trajectory with low specific angular momentum $j<\jiso\simeq4r_{g}c$,
through the innermost stable periapse distance $r_{\mathrm{ISO}}$.
It is this angular momentum barrier, rather than the Eddington limit,
that is the main obstacle to supra-exponential growth.

The BH is more massive than a cluster star, so $\vbh^{2}<\cinf^{2}$,
and the accretion flow is quasi-spherical. In the idealized case where
the flow is radial and adiabatic, it is described by the Bondi solution
\cite{M}{bon52}, $\dot{M}_{B}=(\pi/\sqrt{2})r_{a}^{2}\dinf\cinf$
(adiabatic index $\Gamma=4/3$ assumed), which can be written compactly
in terms of $\mu=\Mbh/M_{i}$, where $M_{i}$ is the initial BH mass,
as $\dot{\mu}=\mu^{2}/t_{B}$, with timescale $t_{B}=\cinf^{3}\left/2^{3/2}\pi G^{2}M_{i}\dinf\right.$.
The stronger-than-linear dependence of the accretion rate on the BH
mass leads to a solution that diverges supra-exponentially in a finite
time $t_{B}$ as $\mu(t)=1/(1-t/t_{B})$. Physical flows, where gravitational
energy is released as radiation, are not strictly adiabatic. As the
mass accretion rate grows, the local luminosity can far exceed the
Eddington luminosity $L_{E}=4\pi cG\Mbh/\kappa=\dot{M}_{E}c^{2}$
($\kappa$ is the gas opacity), for which radiation flux pressure
balances gravity. However, radiation produced inside the photon-trapping
radius $\rtr\sim(\Mdot/\dot{M}_{E})r_{g}$ is carried with the flow
into the BH, because the local optical depth $\tau(r)\sim\kappa\rho(r)r$
makes photon diffusion outward slower than accretion inward \cite{M}{beg78}
{[}which is a manifestation of the ${\cal O}(v/c)$ effect of relativistic
beaming \cite{M}{mih+84}{]}. The luminosity $L_{\infty}$ that
escapes to infinity from $r\gtrsim r_{\gamma}$ translates to a lowered
radiative efficiency $\eta_{\gamma}=L_{\infty}/\dot{M}c^{2}\sim\min(r_{g}/r_{\gamma},r_{g}/r_{\mathrm{ISO}})$,
so it does not exceed $\sim L_{E}$, thereby allowing supra-exponential
Bondi mass accretion rates \cite{M}{sof82}. Detailed calculations
show that $L_{\infty}\lesssim0.6L_{E}$ \cite{M}{beg79}. The associated
radiation pressure enters the dynamics of the flow as an effective
reduction of gravity by a factor $0.4$, and consequently, as a reduction
of the accretion rate by $0.4^{2}\simeq1/6$. The supra-exponential
divergence of spherical accretion is therefore
\begin{equation}
\Mbh(t)=\frac{M_{i}}{1-t/t_{\infty}}\,,\,\,\,\tinf\simeq6t_{B}=\frac{3}{2^{1/2}\pi}\frac{\cinf^{3}}{G^{2}M_{i}\rho_{\infty}}\,.\label{e:M.tinf}
\end{equation}

The typical lifetime of cold flow streams seen in simulations is $\gtrsim10^{7}$
years \cite{M}{wis+08}, which is long enough for the $\sim10^{5}\,\Mo$
of gas in the marginally unstable reservoir on the few parsec scale
to accrete onto the growing BH, and yet short enough to be relevant
for forming $z>7$ quasars. As a demonstration of concept, we adopt
the gas properties found in these simulations, and match a divergence
time of $\tinf\sim\mathrm{few\times}10^{7}$ years to the mean physical
parameters of the $4\times10^{4}\,\Mo$ of gas on the $0.25$-pc scale.
We further assume that half of that mass is in a star cluster with
$1\,\Mo$ stars in a nonsingular distribution (a Plummer law), containing
a $10\,\Mo$ stellar BH, and that the escaped radiation from $r>\rtr$
does not substantially affect the cold flow on larger scales. This
cluster, while dense, is dynamically stable on timescales $\gg\tinf$.
Table \ref{t:M.model} lists the input physical parameters of the
model, the derived gas and cluster properties, and the predicted accretion
properties. 

\begin{table}[!h]
\begin{centering}
\protect\caption{\label{t:M.model}Supra-exponential BH growth model}

\par\end{centering}

\centering{}%
\begin{tabular}{lcc}
\multicolumn{3}{l}{}\tabularnewline
\hline 
\textbf{Property$^{\star}$}  & \textbf{Notation}  & \textbf{Value}\tabularnewline
\hline 
\hline 
\multicolumn{3}{c}{\emph{Model parameters}}\tabularnewline
\hline 
Initial BH mass  & $\Mbh$  & $10\,\Mo$\tabularnewline
Star mass  & $\Ms$  & $1\,\Mo$\tabularnewline
Core radius  & $R_{c}$  & $0.25\,\mathrm{pc}$\tabularnewline
Stellar core mass  & $M_{s}$  & $2\times10^{4}\,\Mo$\tabularnewline
Gas mass in core & $M_{g}$  & $2\times10^{4}\,\Mo$\tabularnewline
Total core mass  & $M_{c}$  & $4\times10^{4}\,\Mo$\tabularnewline
Cold flow adiabatic index & $\Gamma$  & $5/3$\tabularnewline
\hline 
\multicolumn{3}{c}{\emph{Derived cluster properties}}\tabularnewline
\hline 
Mean gas density  & $\dinf$  & $2.1\times10^{-17}\,\mathrm{g\, cm^{-3}}$\tabularnewline
Jeans sound speed  & $c_{s}$  & $14.5\,\mathrm{km\, s^{-1}}$\tabularnewline
Velocity dispersion  & $\sigs$  & $18.0\,\mathrm{km\, s^{-1}}$\tabularnewline
Initial BH rms velocity  & $\vbh$  & $11.4\,\mathrm{km\, s^{-1}}$\tabularnewline
Initial BH rms scattering distance  & $\dbh$  & $6.5\times10^{-2}\,\mathrm{pc}$ \tabularnewline
Cluster orbital frequency  & $\Omega_{0}$  & $1/5545\,\mathrm{year^{-1}}$\tabularnewline
Central 2-body relaxation time  & $t_{r0}$  & $1.4\times10^{7}\,\mathrm{years}$\tabularnewline
Vector resonant relaxation time & $t_{vRR}$ & $6.2\times10^{5}\,\mathrm{years}$\tabularnewline
Evaporation time  & $t_{\mathrm{evap}}$  & $8.0\times10^{9}\,\mathrm{years}$\tabularnewline
Collisional destruction time  & $t_{\mathrm{coll}}$  & $1.5\times10^{11}\,\mathrm{years}$\tabularnewline
Gas reservoir dynamical time & $t_{\mathrm{res}}$ & $\sim10^{6}\,\mathrm{years}$\tabularnewline
\hline 
\multicolumn{3}{c}{\emph{Predicted accretion properties}}\tabularnewline
\hline 
Initial accretion radius  & $r_{a}^{0}$  & $2.5\times10^{-4}\,\mathrm{pc}$\tabularnewline
Mass divergence time  & $\tinf$  & $3.5\times10^{7}\,\mathrm{years}$\tabularnewline
Initial specific accretion ang. mom. & $j_{a}/\jiso$ & $1.6$\tabularnewline
Ang. mom. suppression by resonant relaxation & $\sqrt{\tinf/t_{vRR}}$ & $7.6$\tabularnewline
\hline 
\multicolumn{3}{l}{{\footnotesize{}$^{\star}\,$See also definitions and discussion in
the Supplementary Materials.}}\tabularnewline
\hline 
\end{tabular}
\end{table}

Simulations \cite{M}{wis+08} find that the cold flow is nearly
pressure-supported, and thus has little angular momentum. It is also
plausible that there is little net rotation between the gas and the
stars that were formed from it. However, unavoidable gravitational
interactions of the low-mass BH with cluster stars accelerate it at
orbital frequencies of $\Omega(\rbh)\sim\Omega_{0}$, which induce
a velocity gradient across the capture radius. Gas captured by the
BH then acquires specific angular momentum relative to it, $j_{a}$,
due to the opposite velocity and density gradients and the velocity-skewed
capture cross-section ($j_{a,v}\sim\Omega r_{a}^{2}$ and $j_{a,\rho}\sim(\mathrm{d\log}\rho/\mathrm{d}\log r)_{\rbh}\Omega r_{a}^{2}$)
\cite{M}{ruf+95}. Gas with $j_{a}>\jiso$ cannot plunge directly
into the BH but rather circularizes at a radius $r_{c}=j_{a}^{2}/G\Mbh$,
and accretion then proceeds on the slow viscous timescale rather than
the fast, near-free fall timescale. Analytic and numeric results on
the capture efficiency of angular momentum by inhomogeneous wind accretion
are currently available only in the ballistic (supersonic) limit \cite{M}{liv+86,ruf99}.
We adapt these results to provide a rough estimate of the angular
momentum in the accretion flow on the BH seed in the subsonic regime
considered here (see details in the Supplementary Materials). 

The formal divergence of $j_{a}/\jiso\propto\Mbh$ in the test particle
limit, where $\Omega\sim\Omega_{0}$, is reversed by the accretion
itself. The nonrotating gas accreted by the BH from the cluster exerts
a drag on it, $\ddot{\boldsymbol{r}}_{\bullet}=-2\gamma_{a}\boldsymbol{\dot{r}_{\bullet}}$,
where $\gamma_{a}=\Mdot/2\Mbh=\Mbh(t)/2M_{i}\tinf$. This provides
additional damping beyond dissipation by dynamical friction against
the gas and stars, $\gamma_{\mathrm{df}}\propto[\Mbh(t)/M_{c}]\Omega_{0}$
\cite{M}{bin+08}, which is balanced by the two-body fluctuations.
Accretion damping drives the BH to subequipartition energy and angular
momentum. Two-body interactions with the cluster stars can reestablish
equipartition on the central relaxation timescale $t_{r0}\sim2\times10^{7}$
years only as long as the BH growth rate $2\gamma_{a}$ is slower
than the relaxation rate, up to time $t_{\mathrm{eq}}/\tinf=1-t_{r0}/\tinf\sim0.6$
(Eq. \ref{e:M.tinf}), when the BH has grown by a factor of only $\tinf/t_{r0}\sim2.5$
to $M_{\mathrm{eq}}\simeq25\,\Mo$. At later times, $j_{a}$ is expected
to fall below the extrapolated test particle limit, because both the
BH wandering radius $\dbh$ and the orbital frequency $\Omega$ are
increasingly damped by dynamical friction and by the accretion drag,
which are both $\propto\Mbh(t)$.

\begin{figure}[!p]
\noindent \begin{centering}
\includegraphics[clip,width=1\columnwidth]{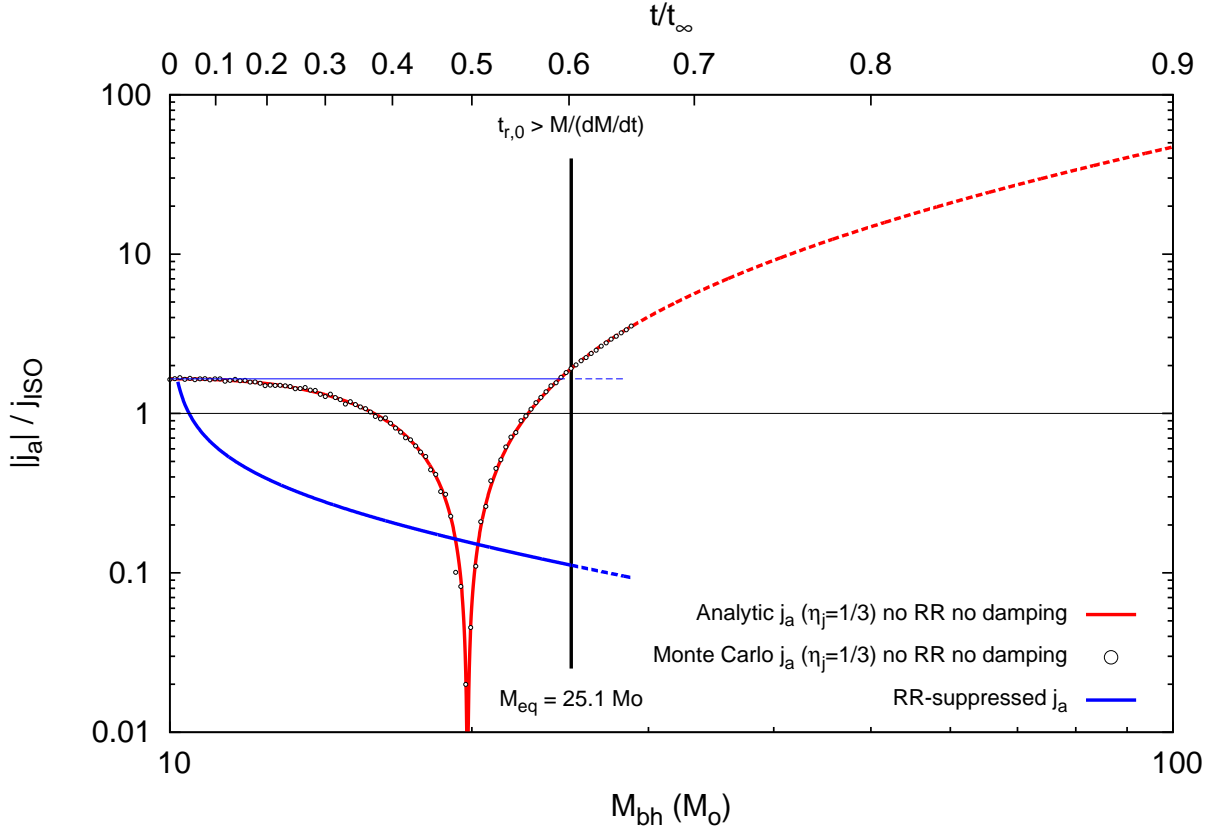}
\par\end{centering}

\protect\caption{\label{f:M.ja}The specific angular momentum ratio, $j_{a}/j_{\mathrm{ISO}}$,
in gas captured by the BH, as function of the BH mass $\protect\Mbh$
(and the corresponding $t/\protect\tinf$ for Bondi accretion). 
The evolution of $j_{a}/j_{\mathrm{ISO}}$ during the initial stages
of the BH growth is calculated in the ballistic wind accretion limit
by a first-order analytic estimate with an angular momentum capture
efficiency of $\eta_{j}=1/3$ (red line), which is validated against
results from a Monte Carlo integration over the exact capture cross-section
(circles). $j_{a}$ falls to zero at $M_{0}\simeq20\,\protect\Mo$,
where the density and velocity gradients cancel each other. The vertical
line at $M_{\mathrm{eq}}\simeq25\,\protect\Mo$ marks the transition
to a dynamical regime where two-body relaxation can no longer establish
equipartition between the growing BH and the stars, and the acceleration
frequency $\Omega$ is damped. The early dynamical suppression of
the angular momentum down to $j_{a}<j_{\mathrm{ISO}}$ by resonant
relaxation of the BH orbit (thick blue line) is approximated by conservatively
assuming a constant $j_{a}(t)=j_{a}(0)$ (thin blue line). At $\protect\Mbh>M_{\mathrm{eq}}$,
where the BH dynamics are sub-equipartition, the actual value of $j_{a}/\protect\jiso$
is expected to progressively drop below the extrapolated one (dashed
lines). (See the detailed discussion in the Supplementary Materials).}
\end{figure}

Figure (\ref{f:M.ja}) shows the evolution of $j_{a}/\jiso$ with
$\Mbh$ for the cold flow cluster model of table (\ref{t:M.model}).
A key property of acceleration-induced angular momentum accretion
is the existence of a BH mass scale $M_{0}$ where $j_{a}(M_{0})\to0$
on typical orbits, because the density and velocity gradients cancel
each other. For a cluster in dynamic equipartition and pressure-supported
gas, $M_{0}$ depends weakly only on the shape of the density/potential
cluster model near the origin: For the Plummer model, $M_{0}\simeq20\Ms$.
This low mass scale is significant because $M_{i}<M_{0}<M_{\mathrm{eq}}$,
and therefore $j_{a}/\jiso$ remains low during the critical stage
of early growth, before damping can become effective. 

In addition, vector resonant relaxation \cite{M}{rau+96,hop+06a},
a rapid process of angular momentum relaxation that operates in nearly
spherical potentials, further suppresses the growth of $j_{a}$ by
randomizing the BH's orbital orientation on a timescale\cite{M}{footnote2}
$t_{vRR}\sim6\times10^{5}$ years. This decreases $j_{a}$ by a factor
of $\sqrt{\tinf/t_{vRR}}\sim8$ over the divergence time. Randomization
by resonant relaxation can be effective until time $t_{\mathrm{rnd}}/\tinf=1-t_{vRR}/\tinf\sim0.98$.
By that time, the BH has grown by a factor of $\tinf/t_{vRR}\sim60$
to $M_{\mathrm{rnd}}\simeq600\,\Mo$ and $>0.95$ of its mass has
been accreted from low-angular momentum gas in the absence of efficient
equipartition. The effect of resonant relaxation can be estimated
analytically in the early growth stages, up to $\Mbh\sim M_{\mathrm{eq}}$,
when both the growth rate and $j_{a}$ can be approximated as near-constant:
Dynamical randomization quickly pushes $j_{a}/\jiso$ below $1$ (independently
of the decrease in $j_{a}/\jiso$ near $\Mbh=M_{0}$) (Fig. \ref{f:M.ja}).
The damped and randomized motions of the BH suppress the accumulation
of angular momentum in the accretion flow and allow Bondi accretion
to proceed supra-exponentially.

The BH mass up to $t_{\mathrm{rnd}}$, $\Mbh\le M_{\mathrm{rnd}}\sim0.02M_{c}$
is still small enough to justify both treating the BH as a test particle
in the fixed potential of the gas and star cluster and representing
cluster dynamics by a simple model. It is much more difficult to self-consistently
predict the subsequent joint evolution of the BH and cluster. However,
the physical arguments for the gradual deceleration of the BH and
the decline of $j_{a}$ beyond $M_{\mathrm{eq}}\sim25\,\Mo$, suggest
that a substantial fraction of the available $10^{5}\,\Mo$ gas reservoir
can be accreted in $\tinf\sim\mathrm{few\times}10^{7}$ years at $z>15$.
Even if the supra-exponential growth phase terminates with a modest
BH seed of only $M_{\mathrm{rnd}}=600\,\Mo$ at $z=16$ ($t\simeq0.25$
billion years for $H_{0}=0.7$, $\Omega=1$, $\Omega_{M}=0.28$),
this would allow the subsequent Eddington-limited growth (with radiative
efficiency $\eta_{\gamma}=0.1$ and electron-scattering opacity $\kappa=0.35\,\mathrm{cm^{2}}\,\mathrm{g^{-1}}$)
of a $3.4\times10^{8}\,\Mo$ SMBH by $z=7$ ($t\simeq0.78$ billion
years), and a $2.5\times10^{10}\,\Mo$ one by $z=6$ ($t\simeq0.95$
billion years). Even if the process operates efficiently only in 1
to 5\% of the dark matter halos where the first stars form, it can
adequately account for the SMBHs seen to be powering the detected
luminous quasars at $z>6$.

We conclude that low mass stellar BHs in very dense, low-angular momentum
cold flows at redshifts $z>15$ can be launched by stellar dynamical
processes into a phase of supply-limited, supra-exponential accretion
and can grow rapidly in $\sim\mathrm{few\times}10^{7}$ years into
$\gtrsim10^{4}\,\Mo$ BH seeds. Subsequent slower Eddington-limited
growth by disk accretion suffices to produce the supermassive BHs
that power the brightest early quasars. 

\chead{}\bibliography{M}{M,footnotes}{References and Notes}

\subsection*{Acknowledgments}

We thank P. Armitage, B. Bar-Or, M. Begelman, F. Bournaud, M. Colpi,
A. Dekel. J.-P. Lasota, C. Reynolds and N. Sapir for helpful discussions
and comments. T.A. acknowledges support by European Research Council
Starting Grant No. 202996, DIP-BMBF Grant No. 71-0460-0101, and the
I-CORE Program of the PBC and Israel Science Fund (Center No. 1829/12).
P.N. acknowledges support from a NASA-NSF Theoretical and Computational
Astrophysics Networks award number 1332858. The authors thank the
Kavli Institute for Theoretical Physics, UC Santa Barbara, where this
work was initiated and supported in part by NSF Grant PHY11-2591.
T.A. is grateful for the warm hospitality of Angel Mill\'{a}n and
Lucy Arkwright of La Posada San Marcos, Al\'{a}jar, Spain, who hosted
the Al\'{a}jar Workshop where this work was continued.

\newpage{}

\pagebreak{}

\pagestyle{fancyplain}

\begin{centering}

\rhead{}\lhead{}\chead{T. Alexander, P. Natarajan, Science, 7 August 2014 (10.1126/science.1251053).}

\section*{Supplementary Materials for}

\section*{Rapid Growth of Seed Black Holes in the Early Universe\protect \\
by Supra-Exponential Accretion}

\rule{0em}{2em}Tal Alexander$^{1}$ and Priyamvada Natarajan$^{2}$\\
{\scriptsize{}$^{1}$Department of Particle Physics \& Astrophysics,
Weizmann Institute of Science, Rehovot 76100, Israel}\\
\rule{0em}{1em}{\scriptsize{}$^{2}$Department of Astronomy, Yale
University, 260 Whitney Avenue, New Haven, CT 06511, USA}{\scriptsize \par}

\rule{0em}{2em}correspondence to: tal.alexander@weizmann.ac.il

\end{centering}

\rule{0em}{4em}\textbf{This PDF file includes:}

\rule{4em}{0em}\rule{0em}{2em}Supplementary Text

\rule{4em}{0em}Fig. S1

\rule{4em}{0em}References (44--54)

\newpage{}

\setcounter{page}{1}
\setcounter{equation}{0} 
\setcounter{figure}{0}
\setcounter{footnote}{0}
\renewcommand{\theequation}{S\arabic{equation}}
\renewcommand{\thesection}{S\arabic{section}}
\renewcommand{\thesubsection}{S\arabic{subsection}}
\renewcommand{\thefigure}{S\arabic{figure}}

\section*{Supplementary Text}

The supra-exponential accretion scenario for the growth of black hole
(BH) seeds in the early universe ties together processes related to
the stellar dynamics of the cluster with those related to the accretion
flow on the BH. These are discussed in this context and derived here
in detail: the dynamics of the BH in the star cluster in section \ref{s:S.dynamics},
and the angular momentum of the accretion flow in section \ref{s:S.Jflow}.

\subsection{\uline{BH dynamics in the host cluster}}

\label{s:S.dynamics}

The dynamical evolution of the gas-rich host stellar cluster, and
the dynamics of the BH in it, can affect the nature of the accretion
flow onto the BH, and the time available for the BH growth. The initial
low-mass BH rapidly reaches a dynamical fluctuation / dissipation
equilibrium with the stars in the dense cluster via two-body interactions.
If the stellar velocities are Maxwellian, the equilibrium is that
of equipartition. For the Plummer density model\cite{S}{dej87}
that is assumed here for quantitative estimates%
\footnote{Define $M_{\mathrm{tot}}=2^{3/2}M_{c}$, $x=r/R_{c}$, and $y=\sqrt{1+x^{2}}$.
The Plummer density profile is $\rho=(3M_{\mathrm{tot}}/4\pi R_{c}^{3})/y^{5}$,
the enclosed mass is $M_{<}=M_{\mathrm{tot}}x^{3}/y^{3}$, the potential
is $\phi=-(GM_{\mathrm{tot}}/R_{c})/y$, the 1D velocity dispersion
is $\sigma^{2}=-\phi/6$, the fundamental frequency is $\Omega_{0}^{2}=GM_{\mathrm{tot}}/R_{c}^{3}$,
and the azimuthal velocity is $v_{\phi}^{2}=(GM_{\mathrm{tot}}/R_{c})x^{2}/y^{3}$. %
}, equipartition is a good approximation\cite{S}{cha+02b}.

We simplify here the qualitative discussion of the dynamics by approximating
the mass distribution of the nuclear star cluster as a constant density
spherical core of radius $R_{c}$ with mass $M_{c}$ in stars and
gas, and the velocity distribution as Maxwellian with 1D velocity
dispersion $\sigs$. The cluster contains $\Ns$ stars of mass $\Ms$
each (the total stellar mass in the core is $M_{s}=\Ns\Ms\equiv sM_{c}$,
where $s\le1$), and one BH of mass $\Mbh$. We further assume that
$M_{s}/M_{g}\sim{\cal O}(1)$ ($s\sim1/2)$ in the core, and therefore
the stellar dynamics can be approximated, to within order unity corrections,
by neglecting the gas. Conversely, since the cold flows provide an
effectively infinite reservoir of gas compared to $M_{c}$, we assume
that the gas properties are not substantially affected by the stellar
dynamics.

We denote the mass ratio $Q=\Mbh/\Ms$, the core velocity $V_{c}^{2}=GM_{c}/R_{c}$
and the core orbital frequency $\Omega_{c}^{2}=R_{c}^{3}/GM_{c}$.
In equipartition, the mean energy of the BH is $E_{\bullet}=3\Ms\sigs^{2}$,
 and therefore the BH's rms 3D velocity is $\vbh^{2}=3\sbh^{2}=(3/Q)\sigs^{2}$
and its rms 3D displacement from the center is $\dbh^{2}=(3/Q)\sigs^{2}/\Omega_{c}^{2}$.
Since $\sigs^{2}\sim V_{c}^{2}$, it then follows that $\vbh^{2}\sim V_{c}^{2}/Q$
and $\dbh^{2}\sim R_{c}^{2}/Q$ that is, the BH's typical velocity
is significantly smaller than the typical velocity of stars in the
cluster's potential, and its excursions away from the center are confined
to the central regions of the core, where the density is nearly constant.
An exact treatment of the fluctuation/dissipation equilibrium in the
Plummer potential\cite{S}{cha+02b} yields 
\begin{equation}
\vbh^{2}=(2^{5/2}/3)V_{c}^{2}/Q\,,\,\,\,\dbh^{2}=(2/3)R_{c}^{2}/Q.\label{e:S.vbhdbh}
\end{equation}

Two-body scattering by the cluster stars changes the BH's orbital
energy and angular momentum by order unity around their equipartition
values on the relaxation timescale, 
\begin{equation}
t_{r}(r)=0.34\sigma^{3}(r)/G^{2}n_{\star}(r)\Ms^{2}\log\Lambda\,,\label{e:S.tr}
\end{equation}
where $n_{\star}$ is the local stellar number density, the Coulomb
factor is $\Lambda\simeq0.1N_{\mathrm{tot}}$ in a single-mass cluster\cite{S}{gie+94},
and $N_{\mathrm{tot}}$ is the total number of stars in the cluster.
In a realistic cluster with a spectrum of masses, the term $n_{\star}\Ms^{2}$
in the denominator is replaced by $n_{\star}\left\langle \Ms^{2}\right\rangle $,
where $\left\langle \cdots\right\rangle $ denoted an average over
the mass function. Since $\left\langle \Ms\right\rangle ^{2}\le\left\langle \Ms^{2}\right\rangle $
by definition, and typical stellar mass functions have $\left\langle \Ms\right\rangle ^{2}\ll\left\langle \Ms^{2}\right\rangle $,
naively substituting $\Ms$ by $\left\langle \Ms\right\rangle $ in
the single mass expression (Eq. \ref{e:S.tr}) may substantially over-estimate
the actual relaxation time by a factor $\left\langle \Ms^{2}\right\rangle /\left\langle \Ms\right\rangle ^{2}$.
$ $

The lifespan of compact clusters, such as considered here, is limited
by their internal dynamics. The stars in the cluster will eventually
destroy each other by physical collisions. The mean time between grazing
collisions per star is 
\begin{equation}
t_{\mathrm{graze}}^{-1}=16\sqrt{\pi}n_{\star}\sigma\Rs^{2}\left(1+G\Ms/2\sigma^{2}\Rs\right)\,,\label{e:S.tcoll}
\end{equation}
where $\Rs$ is the stellar radius. Full destruction takes multiple
grazing collisions and the typical timescale is $t_{\mathrm{coll}}\sim10t_{\mathrm{graze}}$\cite{S}{mur+91}.
Even if the stars are able to survive collisions, the cluster will
ultimately evaporate on a timescale of $t_{\mathrm{evap}}\sim100t_{rh}$\cite{M}{bin+08},
where $t_{rh}$ is the relaxation time at the half mass radius $r_{h}$
($r_{h}\simeq1.3R_{c}$ for the Plummer model). For the cluster model
considered here, $t_{rh}\simeq0.9\Ns/\left[s^{2}\Omega_{c}\log(0.28\Ns)\right]$.
Cluster dissolution by energetic 2-body ejections of stars is a significantly
slower process, and can be neglected\cite{M}{bin+08}. The two-body
ejection of the BH itself from the cluster is further suppressed by
the mass ratio $1/(1+Q$) and is therefore a low-probability event\cite{S}{heg+03}.

In addition to uncorrelated 2-body relaxation, which is inherent to
any discrete system, stars in potentials with a high degree of symmetry
rapidly randomize their angular momentum by the process of resonant
relaxation (RR) \cite{M}{rau+96,hop+06a}. The central part of
the cluster on length-scale $\dbh$ is expected to be nearly spherically
symmetric, which implies that the $N_{\bullet}\sim\Ns(\dbh)$ background
stars there will tend to conserve their orbital planes, gradually
tracing rosettes. Averaged over time, the effect of the stars on the
BH can be represented by the residual specific torque $\boldsymbol{\tau}$
that results from the superposed torques by $N_{\bullet}$ randomly
oriented mass annuli, whose magnitude is $\tau=A_{\tau}\sqrt{N_{\star}(A_{\star}\dbh)}G\Ms/\dbh$,
where $A_{\tau}$ and $A_{\star}$ are order unity pre-factors. This
residual torque will change the orientation of the BH orbit coherently
($\propto t$), as long as the orbital planes of the background stars
remain approximately fixed, over a coherence time. The longest possible
coherence time is set by the mutual resonant torques between the background
stars, which ultimately randomize their orbits, and hence $\boldsymbol{\tau}$,
on the self-quenching coherence time $t_{\mathrm{coh}}=A_{\mathrm{coh}}J_{c}/\tau$,
where $A_{\mathrm{coh}}$ is an order unity prefactor, and $J_{c}=\sqrt{GM(<\dbh)\dbh}$
is the circular angular momentum, with $M(<\dbh)$ the total mass
inside $\dbh$. On timescales longer than $t_{\mathrm{coh}}$, the
large change over a coherence time, $\tau t_{\mathrm{coh}}$, becomes
the step-size of a random walk evolution $(\propto\sqrt{t}$). The
RR timescale is then defined by $(\tau t_{\mathrm{coh}})\sqrt{t_{RR}/t_{\mathrm{coh}}}=J_{c}$.

For the purpose of randomizing the accretion flow's angular momentum,
the relevant changes are those of the orbital orientation (direction
of the angular momentum vector), and the corresponding coherence timescale
is the self-quenching timescale. This restricted form of RR is known
as vector RR \cite{M}{rau+96}. The values of the numeric pre-factors
for vector RR, $A_{\tau}$ (torque strength), $A_{\star}$ (size of
the effective torquing volume) and $A_{\mathrm{coh}}$ (length of
the coherence time), can be determined by simulations \cite{S}{gur+07,eil+09},
but are poorly known at this time for the configuration of interest
here (a spherical stellar system without a central massive BH). Based
on the available results, we conservatively estimate 
\begin{equation}
t_{vRR}(\dbh)\sim3\sqrt{\frac{M_{c}}{s\Ms}}\Omega_{c}^{-1}\left(\frac{\dbh}{R_{c}}\right)^{3/2}\,,\label{e:S.tvRR}
\end{equation}
As in the case of 2-body relaxation, a stellar mass spectrum typically
accelerates RR. For vector RR, the substitution of $\Ms$ by $\left\langle \Ms\right\rangle $
over-estimates the relaxation time by a factor of $\left\langle \Ms^{2}\right\rangle ^{1/2}/\left\langle \Ms\right\rangle $.

The cluster potential fixes the Jeans scale for gravitational instability.
Simulations \cite{M}{wis+08,bou+11} indicate that the cold flows
create an isothermal cusp that is nearly pressure supported, implying
that the gas mass inside the Jeans length $\lambda_{J}\sim\sqrt{\pi c_{s}^{2}/G\bar{\rho}_{c}}$,
($\bar{\rho}_{c}=3M_{c}/4\pi R_{c}^{3}=\bar{\rho}_{g}/(1-s)$) is
close to the Jeans mass \cite{M}{bin+08},
\begin{equation}
M_{g,J}\sim\frac{4\pi}{3}\left(\frac{\lambda_{J}}{2}\right)^{3}\bar{\rho}_{g}=\frac{\pi^{5/2}}{6}\frac{c_{s}^{3}}{G^{3/2}\bar{\rho}_{c}^{1/2}}(1-s)\,.\label{e:S.MJ}
\end{equation}
Marginal stability then implies that the sound speed is 
\begin{equation}
c_{s}^{2}\simeq\frac{3}{\pi^{2}}V_{c}^{2}=\frac{3}{\pi^{2}}\frac{GM_{c}}{R_{c}}\,.\label{e:S.cs}
\end{equation}

The accretion of non-rotating, pressure-supported gas by the BH exerts
a drag force on it, $\ddot{\boldsymbol{r}}_{\bullet}=-(\Mdot/\Mbh)\boldsymbol{\dot{r}_{\bullet}}$.
When the mass accretion rate rises above the 2-body relaxation rate,
equipartition can no longer be maintained, and the BH dynamics can
be approximately described as those of a damped 3D harmonic oscillator
in the potential of the constant density core,
\[
\ddot{\boldsymbol{r}}_{\bullet}=-2(\gamma_{a}+\gamma_{\mathrm{df}})\dot{\boldsymbol{r}}_{\bullet}-\Omega_{0}^{2}\boldsymbol{r}_{\bullet}\,,
\]
where $\gamma_{a}(t)=\Mdot(t)/2\Mbh=\Mbh(t)/2M_{i}\tinf$ (for the
Bondi solution) is the accretion damping coefficient, and $\gamma_{\mathrm{df}}(t)=(2^{3/2}\pi^{1/2}/3)\log\Lambda G^{2}\rho_{c}\Mbh(t)/\sigma^{3}$
is the dynamical friction damping coefficient \cite{M}{bin+08}.
Therefore, after the BH decouples dynamically from the stellar cluster,
the amplitude of the orbital oscillations decays, and since both damping
coefficients scale as $\propto\Mbh$, so does the orbital frequency.

\[
\]

\subsection{\uline{Angular momentum in the accretion flow}}

\label{s:S.Jflow}

\begin{figure}
\noindent \centering{}\includegraphics[clip,width=0.7\columnwidth]{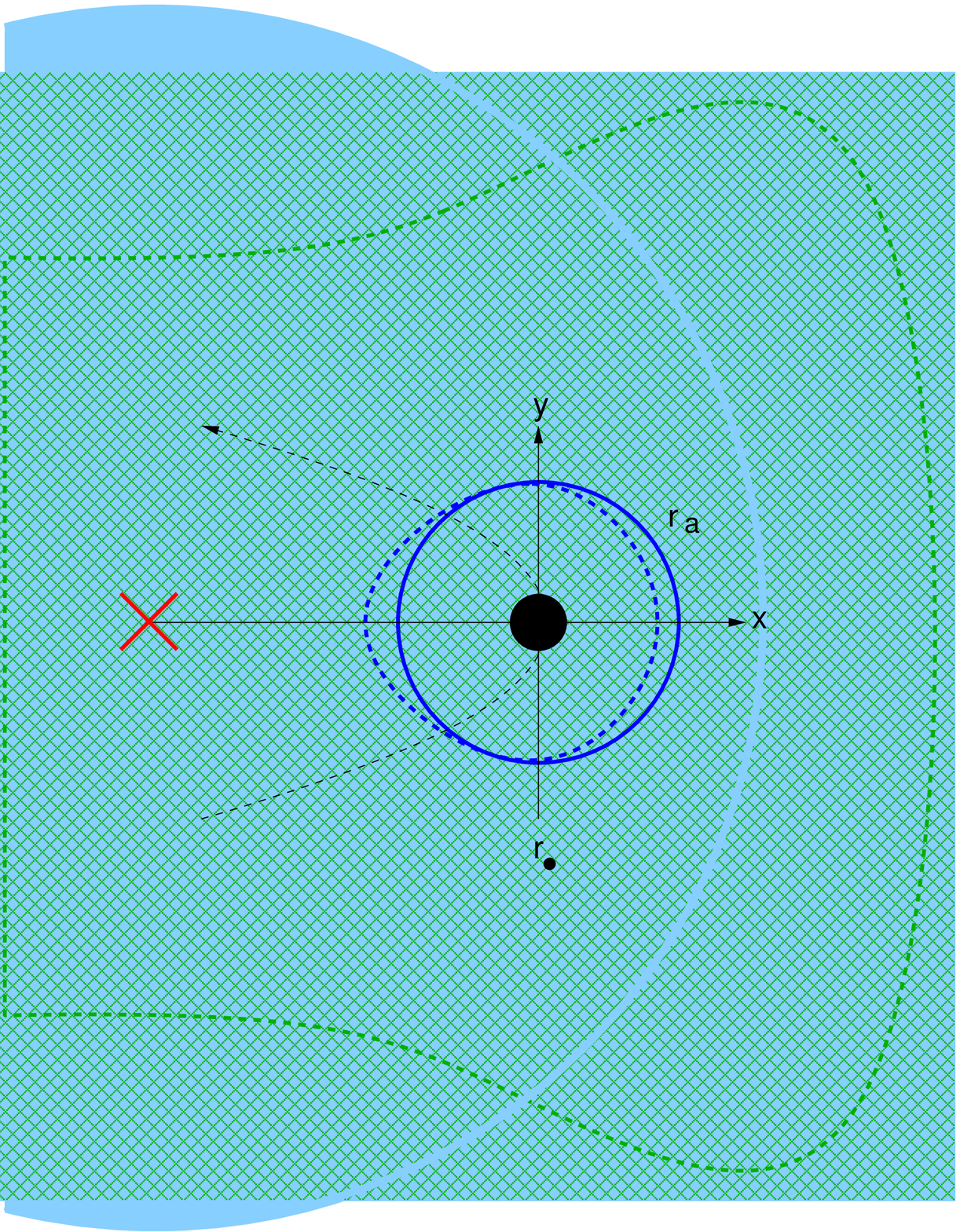}\protect\caption{\label{f:S.dvdr}The geometry assumed for estimating the angular momentum
of the accreted wind in the ballistic approximation. A BH (black circle)
accretes from the gas reservoir (green hashed area) as it orbits (dashed
arrow) at radius $\protect\rbh$ around the center (red cross) of
the stellar core (light blue area). The accretion cross-section of
mean radius $r_{a}(\protect\rbh)$ (dark blue circle) intercepts the
flow on either side of the $y$-axis. The properties of the captured
gas are asymmetric relative to the $y$-axis: the density falls with
$x$ as a Plummer law, while the induced velocity $v(\protect\rbh+x)=\Omega(\protect\rbh)(\protect\rbh+x)$
rises with $x$. The velocity gradient also skews the accretion radius
(blue dashed curve). As a result, the captured gas in the accretion
flow acquires angular momentum relative to the BH (see analysis in
Section \ref{s:S.Jflow}). }
\end{figure}

A basic ingredient of this supra-exponential growth scenario is that
the initial BH is a low-mass stellar BH, which is therefore scattered
substantially by the random perturbations of the cluster stars. This
results in accelerated motion, which in turn induces angular momentum
in the captured gas relative to the accreting BH, due to the velocity
and density gradients across the capture cross-section (Figure \ref{f:S.dvdr};
\cite{S}{fra+02}). This occurs even though the gas in the cluster
is pressure-supported, and therefore not rotating relative to the
cluster center, and the stellar system, which is assumed to have formed
from the gas, is not rotating as a whole relative to the gas. A necessary
condition for prompt accretion is that the specific angular momentum
in the captured wind, $j_{a}$, be lower than that of a plunge orbit,
$\jiso=4r_{g}c$ (parabolic orbit assumed). This allows the gas to
flow directly into the wandering BH. Otherwise, it settles into a
slowly-draining viscous accretion disk at the circularization radius
$r_{c}\simeq j_{a}^{2}/G\Mbh$. This constitutes an angular momentum
barrier that slows down the growth of the BH \cite{S}{ill+00}.

We focus here on the early stages of the BH growth, when the BH mass
is still low enough for it to wander substantially away from the center
at a large fraction of the sound speed, and when its accretion radius
is still much smaller than its wandering radius ($r_{a}\ll\dbh,$
which corresponds to $\Mbh\ll300\,\Mo$ for the model of table \ref{t:M.model}),
so the accretion can be described in terms of a wind. However, this
regime of subsonic wind accretion, and in particular the question
of angular momentum capture by motion relative to an inhomogeneous
medium, is little explored and poorly understood. Analytic and numeric
results on the capture efficiency of angular momentum from an inhomogeneous
wind are currently available only in the ballistic (supersonic) limit
\cite{M}{liv+86,ruf99}. We adopt these results here and modify them
to provide a rough estimate of the angular momentum accreted by the
wandering BH seed. More detailed work, and in particular numeric simulations,
are required for validating this analysis.

Ballistic wind accretion, which is a reasonable approximation in the
hypersonic limit where gas pressure can be neglected, occurs when
initially parallel flow lines in the accretion cylinder on diametrically
opposed sides of the accretor, are focused behind it, intersect dissipatively
by shock, cancel their transverse momentum, become bound to the accretor,
and finally fall radially on the BH from the back. When the wind is
homogeneous, the line of intersection is on the axis of symmetry.
Davies and Pringle \cite{S}{dav+80} showed that density or velocity
gradients in a 2D (planar) flow do not change the outcome, to 1st
order in the gradient: both the mass accretion rate and the zero angular
momentum in it remain as they were in the homogeneous case. Instead,
the intersection line curves to compensate for the imbalance in the
transverse momentum, and the shocked gas then falls in radially from
the point of intersection. However, Davies and Pringle noted that
these results can not be directly carried over to 3D flows. The 3D
case was addressed by hydrodynamic simulations. Livio, Soker and collaborators
\cite{M}{liv+86} studied flows with density gradients, and found
that the mass capture rate remains close to that of homogeneous wind
accretion, but the fraction of specific angular momentum that is captured
is only $\eta_{j}\sim0.10$--$0.25$ of that available in the accretion
cylinder. Ruffert studied flows with both density \cite{M}{ruf99}
and velocity \cite{S}{ruf97} inhomogeneities with higher resolution
simulations, and found that the mass capture efficiency remains high,
and the mean fraction of captured angular momentum varies with very
strong dependence on the flow parameters, $\eta_{j}\sim0.0$--$0.7$.
We therefore adopt here a representative angular momentum capture
efficiency of $\eta_{j}=1/3$. 

A density gradient across the capture cross-section is present when
the velocity has a transverse component relative to the radius vector,
and an induced velocity gradient is present when the acceleration
has a transverse component relative to the velocity vector. The gradients
are therefore maximal on a circular orbit. For an isotropic distribution
of velocities and accelerations, the rms gradients are $\sqrt{2/3}$
times smaller. Consider for simplicity a BH on a circular orbit at
radius $\rbh$ from the cluster center, moving with velocity $v_{0}(\rbh)=\Omega(\rbh)\rbh$
relative to gas with density $\rho_{0}(\rbh)$. Denote by $x$ the
location relative to the BH along the radial direction (i.e. along
the gradients) in the plane of the capture cross-section, and by $y$
the location in the direction perpendicular to it and to the acceleration
vector (Figure \ref{f:S.dvdr}). A velocity gradient also affects
the accretion of angular momentum by modifying the size of the accretion
radius, $r_{a}(x)=2G\Mbh/[c_{s}^{2}+v^{2}(\rbh+x)]$, thereby skewing
the accretion cross-section: it is more extended where the induced
velocity is smaller, on the side closer to the cluster center, and
less extended on the opposite side. It is convenient to measure velocities
in terms of the asymptotic sound speed, $u=v/c_{s}$, and distances
in the $x-y$ plane in terms of the unperturbed accretion radius $r_{a}(0)=2G\Mbh/(1+u_{0}^{2})$,
so that $d_{a}(x)=r_{a}(x)/r_{a}(0)=(1+u_{0}^{2})/(1+u^{2}(x))$.
The variations in density and velocity along the $x$-axis can be
described to 1st order by the dimensionless gradients $\epsr$, $\epsu$,
\begin{equation}
\rho(x)=\rho_{0}(1+\epsr x)\,,\qquad u(x)=u_{0}(1+\epsu x)\,.
\end{equation}
Assuming that the cluster gas is pressure-supported ($c_{s}^{2}=(3/\pi^{2})V_{c}^{2}$),
the typical scale of the gradients here is $\epsilon\sim{\cal O}(r_{a}/R_{c})=(2\pi^{2}/3)(\Mbh/M_{c})\sim10^{-3}$.

Gas approaching the BH with impact parameters ($x,y$) is captured
when $\sqrt{x^{2}+y^{2}}<d_{a}(x)$. The mass capture rate through
an element $\mathrm{d}x\mathrm{d}y$ of the accretion cross-section
is $\mathrm{d}\dot{M}_{a}=\rho(x)v(x)\mathrm{d}x\mathrm{d}y$, and
the angular momentum capture rate is $\mathrm{d}\dot{J}_{a}=\rho(x)v^{2}(x)x\mathrm{d}x\mathrm{d}y$.
The total accretion rate through the capture cross-section is therefore
\cite{M}{ruf+95} 
\begin{eqnarray}
\dot{M}_{a} & = & 2\int_{x_{-}(\epsilon_{u})}^{x_{+}(\epsilon_{u})}\rho(x)u(x)\sqrt{d_{a}^{2}(x)-x^{2}}\mathrm{d}x\simeq\pi\rho_{0}u_{0}\,,\nonumber \\
\dot{J}_{a} & = & 2\int_{x_{-}(\epsilon_{u})}^{x_{+}(\epsilon_{u})}\rho(x)u^{2}(x)x\sqrt{d_{a}^{2}(x)-x^{2}}\mathrm{d}x\simeq\frac{1}{4}\left[\epsr-2\frac{(3u_{0}^{2}-1)}{u_{0}^{2}+1}\epsu\right]\pi\rho_{0}u_{0}^{2}\,,
\end{eqnarray}
where the integration limits $x_{-}<0$ and $x_{+}>0$ are the solutions
of the cubic equation $x_{\pm}=\pm d_{a}(x_{\pm})$, and the approximate
expressions are to first order in $\epsr$ and $\epsu$ for finite
velocities (unlike \cite{M}{ruf+95}, where $u_{0}\to\infty$).
Converting back to dimensional units and reinstating the notation
$r{}_{a}=r_{a}(0)$, the specific angular momentum in the accretion
flow is generally 
\begin{eqnarray}
j_{a} & = & \dot{J}_{a}/\dot{M}_{a}=\frac{1}{4}\left(\epsr-2\left(\frac{3u_{0}^{2}-1}{u_{0}^{2}+1}\right)\epsu\right)v_{0}r_{a}\nonumber \\
 & = & \frac{1}{4}\left(\left.\frac{\mathrm{d}\log\rho}{\mathrm{d}\log r}\right|_{\rbh}-2\left(\frac{3u_{0}^{2}-1}{u_{0}^{2}+1}\right)\left.\frac{\mathrm{d}\log v}{\mathrm{d}\log r}\right|_{\rbh}\right)\Omega(\rbh)r_{a}^{2}\,.
\end{eqnarray}

For the Plummer model assumed here, the dimensionless density gradient
is always negative, 
\begin{equation}
\left.\frac{\mathrm{d}\log\rho}{\mathrm{d}\log r}\right|=-5\frac{(r/R_{c})^{2}}{(1+(r/R_{c})^{2})}\,.
\end{equation}
The induced velocity gradient across the accretion cross-section due
to the accelerated motion of the BH in a non-rotating cluster is purely
geometrical (the projection of the BH's angular velocity on the position
of the accretion radius), $v(\rbh+r_{a})=\Omega(\rbh)(r+r_{a})$,
so $\mathrm{d}\log v/\mathrm{d}\log r=1$. The 1st order estimate
of the accreted specific angular momentum in the ballistic approximation
for finite velocities, averaging of isotropic orbital velocities and
accelerations and taking into account the typical accretion efficiency
$\eta_{j}=1/3$, is therefore

\begin{equation}
j_{a}(\rbh)=\frac{1}{4}\sqrt{\frac{2}{3}}\eta_{j}\left\{ \left.\frac{\mathrm{d}\log\rho}{\mathrm{d}\log r}\right|_{\rbh}-2\left[\frac{3(v_{\phi}(\rbh)/c_{s})^{2}-1}{(v_{\phi}(\rbh)/c_{s})^{2}+1}\right]\right\} \Omega(r_{\bullet})r_{a}^{2}\,.\label{e:S.ja}
\end{equation}

Substituting for $\rbh$ the wandering radius in equipartition, $z_{\bullet}=\dbh/R_{c}=\sqrt{2/3Q}$,
so that $\mathrm{d}\log\rho/\mathrm{d}\log r=-5z_{\bullet}^{2}/(1+z_{\bullet}^{2})$
and assuming that the gas is pressure supported, so that $(v_{\phi}/c_{s})^{2}=(2^{3/2}/3)\pi^{2}z_{\bullet}^{2}/(1+z_{\bullet}^{2})^{3/2}$$ $,
fixes the mass ratio $Q$ for which $j_{a}=0$ to $Q_{0}\simeq19.8$
in the Plummer model. With these assumptions, the value of $Q_{0}$
depends only on the dimensionless functional form of the potential
/ density pair that describes the cluster. Since typically $z_{\bullet}^{2}\ll1$,
its numeric value is not expected to depend strongly on the particular
choice of the non-singular cluster potential. A BH with mass $\Mbh\sim Q_{0}\Ms$
on a typical orbit with $\rbh\sim\dbh$, will therefore tend to accrete
mass with little angular momentum, because the density and velocity
gradients cancel each other (Figure \ref{f:M.ja}). The existence
of this angular momentum minimum near the initial BH mass helps suppress
the accumulation angular momentum in the accretion flow in the critical
initial stages of the BH seed growth.

Cluster dynamics also play such a role by randomizing the BH's orbital
orientation on the short vector resonant relaxation timescale $t_{vRR}$
(see Section \ref{s:S.dynamics}). This continually adds misaligned
gas to the accretion flow, which mixes in with any gas stalled in
an accretion disk, and accelerates its draining into the BH. This
mechanism is particularly important at the early stages of the growth,
when $\Mbh<M_{\mathrm{eq}}=(\tinf/t_{r0})M_{i}\simeq2.5M_{i}$ (at
$t<t_{\mathrm{eq}}=\tinf-t_{r0}\simeq0.6\tinf$), before the BH can
decouple from the perturbing stellar background. We derive a simple
estimate of this suppression, assuming that the BH grows initially
by Bondi accretion, $M(t)=M_{i}/(1-t/\tinf)$, and that $j_{a}$ can
be conservatively approximated as constant, $j_{a}(\Mbh)=j_{a}(M_{i})=j_{i}$,
up to $\Mbh\sim M_{\mathrm{eq}}$ (Figure \ref{f:M.ja}). At these
early times, the growth can be approximated as linear, $M(t)\simeq M_{i}(1+t/\tinf)$.
It then follows that the captured mass grows as $\Delta M=(M_{i}/\tinf)t$,
while the angular momentum in the flow grows in a random walk fashion
as $\Delta J=[j_{i}(M_{i}/\tinf)t_{vRR}]\sqrt{t/t_{vRR}}$ on times
longer than vector RR timescale, $t_{vRR}$. The specific angular
momentum accumulated in the accretion flow therefore falls with time
as 
\begin{equation}
j_{a}(t)=\frac{\Delta J(t)}{\Delta M(t)}\sim j_{i}\sqrt{\frac{t_{vRR}}{t}}\,,\quad(t>t_{vRR})\,,
\end{equation}
or in terms of the BH mass, 
\begin{equation}
j_{a}(\Mbh)=\frac{j_{i}}{\sqrt{N_{vRR}}\sqrt{1-M_{i}/\Mbh}}\,,\quad(\Mbh>\frac{M_{i}}{1-t_{vRR}/\tinf})\,,
\end{equation}
where $N_{vRR}=\tinf/t_{vRR}$ is the number of RR angular momentum
steps over the divergence time ($\sqrt{N_{vRR}}=7.6$ for the cluster
model of table \ref{t:M.model}). Figure \ref{f:M.ja} shows that
early suppression by RR decreases $j_{a}$ rapidly below $\jiso$,
thereby enabling prompt radial accretion until dynamical decoupling
becomes effective. This suppression of angular momentum is further
aided by the angular momentum minimum near $Q_{0}$,

\let\oldthebibliography=\thebibliography 
\let\oldendthebibliography=\endthebibliography 
\renewenvironment{thebibliography}[1]{%
\oldthebibliography{#1}%
\setcounter{enumiv}{43}%
}{\oldendthebibliography} 
\bibliography{S}{S}{References}

\begin{thebibliography}{10}

\bibitem{mor+11}
D.~J. {Mortlock}, {\it et~al.\/}, {\it \nat\/} {\bf 474}, 616 (2011).

\bibitem{fan+06}
X.~{Fan}, {\it et~al.\/}, {\it \aj\/} {\bf 131}, 1203 (2006).

\bibitem{jeo+12}
M.~{Jeon}, {\it et~al.\/}, {\it \apj\/} {\bf 754}, 34 (2012).

\bibitem{mil+09}
M.~{Milosavljevi{\'c}}, V.~{Bromm}, S.~M. {Couch}, S.~P. {Oh}, {\it \apj\/}
  {\bf 698}, 766 (2009).

\bibitem{par+12}
K.~{Park}, M.~{Ricotti}, {\it \apj\/} {\bf 747}, 9 (2012).

\bibitem{vol+05b}
M.~{Volonteri}, M.~J. {Rees}, {\it \apj\/} {\bf 633}, 624 (2005).

\bibitem{hop+07b}
P.~F. {Hopkins}, {\it et~al.\/}, {\it \apj\/} {\bf 662}, 110 (2007).

\bibitem{vol10}
M.~{Volonteri}, {\it \aapr\/} {\bf 18}, 279 (2010).

\bibitem{nat11}
P.~{Natarajan}, {\it Bulletin of the Astronomical Society of India\/} {\bf 39},
  145 (2011).

\bibitem{hai13}
Z.~{Haiman}, {\it Astrophysics and Space Science Library\/}, T.~{Wiklind},
  B.~{Mobasher}, V.~{Bromm}, eds. (2013), vol. 396 of {\it Astrophysics and
  Space Science Library\/}, pp. 293--341.

\bibitem{bro+03}
V.~{Bromm}, A.~{Loeb}, {\it \apj\/} {\bf 596}, 34 (2003).

\bibitem{lod+06}
G.~{Lodato}, P.~{Natarajan}, {\it \mnras\/} {\bf 371}, 1813 (2006).

\bibitem{beg+06}
M.~C. {Begelman}, M.~{Volonteri}, M.~J. {Rees}, {\it \mnras\/} {\bf 370}, 289
  (2006).

\bibitem{lod+07}
G.~{Lodato}, J.~E. {Pringle}, {\it \mnras\/} {\bf 381}, 1287 (2007).

\bibitem{dev+09}
B.~{Devecchi}, M.~{Volonteri}, {\it \apj\/} {\bf 694}, 302 (2009).

\bibitem{dav+11}
M.~B. {Davies}, M.~C. {Miller}, J.~M. {Bellovary}, {\it \apjl\/} {\bf 740}, L42
  (2011).

\bibitem{wil+03}
C.~J. {Willott}, R.~J. {McLure}, M.~J. {Jarvis}, {\it \apjl\/} {\bf 587}, L15
  (2003).

\bibitem{fer+13}
A.~{Ferrara}, F.~{Haardt}, R.~{Salvaterra}, {\it \mnras\/} {\bf 434}, 2600
  (2013).

\bibitem{joh+13}
J.~L. {Johnson}, D.~J. {Whalen}, H.~{Li}, D.~E. {Holz}, {\it \apj\/} {\bf 771},
  116 (2013).

\bibitem{tre+13}
E.~{Treister}, K.~{Schawinski}, M.~{Volonteri}, P.~{Natarajan}, {\it \apj\/}
  {\bf 778}, 130 (2013).

\bibitem{alv+09}
M.~A. {Alvarez}, J.~H. {Wise}, T.~{Abel}, {\it \apjl\/} {\bf 701}, L133 (2009).

\bibitem{gre+11}
T.~H. {Greif}, {\it et~al.\/}, {\it \apj\/} {\bf 737}, 75 (2011).

\bibitem{tur+12}
M.~J. {Turk}, J.~S. {Oishi}, T.~{Abel}, G.~L. {Bryan}, {\it \apj\/} {\bf 745},
  154 (2012).

\bibitem{reg+14}
J.~A. {Regan}, P.~H. {Johansson}, M.~G. {Haehnelt}, {\it \mnras\/} {\bf 439},
  1160 (2014).

\bibitem{saf+14}
C.~{Safranek-Shrader}, M.~{Milosavljevi{\'c}}, V.~{Bromm}, {\it \mnras\/} {\bf
  440}, L76 (2014).

\bibitem{dek+09}
A.~{Dekel}, {\it et~al.\/}, {\it \nat\/} {\bf 457}, 451 (2009).

\bibitem{dub+12}
Y.~{Dubois}, {\it et~al.\/}, {\it \mnras\/} {\bf 423}, 3616 (2012).

\bibitem{bou+11}
F.~{Bournaud}, {\it et~al.\/}, {\it \apjl\/} {\bf 741}, L33 (2011).

\bibitem{wis+08}
J.~H. {Wise}, M.~J. {Turk}, T.~{Abel}, {\it \apj\/} {\bf 682}, 745 (2008).

\bibitem{footnote1}
The near-stability of such a massive reservoir may require suppression of $H_2$
  cooling by strong sources of Lyman-Werner radiation, implying that such
  reservoirs are rare. Neither effect was included explicitly in the
  simulations; however they have been investigated theoretically in detail [for
  example, \cite{M}{aga+13}].

\bibitem{bon52}
H.~{Bondi}, {\it \mnras\/} {\bf 112}, 195 (1952).

\bibitem{beg78}
M.~C. {Begelman}, {\it \mnras\/} {\bf 184}, 53 (1978).

\bibitem{mih+84}
D.~{Mihalas}, B.~W. {Mihalas}, {\it {Foundations of radiation hydrodynamics}\/}
  (1984).

\bibitem{sof82}
M.~H. {Soffel}, {\it \aap\/} {\bf 116}, 111 (1982).

\bibitem{beg79}
M.~C. {Begelman}, {\it \mnras\/} {\bf 187}, 237 (1979).

\bibitem{ruf+95}
M.~{Ruffert}, U.~{Anzer}, {\it \aap\/} {\bf 295}, 108 (1995).

\bibitem{liv+86}
M.~{Livio}, N.~{Soker}, M.~{de Kool}, G.~J. {Savonije}, {\it \mnras\/} {\bf
  222}, 235 (1986).

\bibitem{ruf99}
M.~{Ruffert}, {\it \aap\/} {\bf 346}, 861 (1999).

\bibitem{bin+08}
J.~{Binney}, S.~{Tremaine}, {\it {Galactic Dynamics: Second Edition}\/}
  (Princeton University Press, 2008).

\bibitem{rau+96}
K.~P. {Rauch}, S.~{Tremaine}, {\it New Astronomy\/} {\bf 1}, 149 (1996).

\bibitem{hop+06a}
C.~{Hopman}, T.~{Alexander}, {\it \apj\/} {\bf 645}, 1152 (2006).

\bibitem{footnote2}
Similar randomization by two-body relaxation is negligibly slow, by comparison.
  However, both two-body relaxation and resonant relaxation will be
  substantially accelerated by a realistic stellar mass spectrum, and
  inhomogeneities in the gas flow will also contribute to the randomization of
  the BH orbit.

\bibitem{aga+13}
B.~{Agarwal}, A.~J. {Davis}, S.~{Khochfar}, P.~{Natarajan}, J.~S. {Dunlop},
  {\it \mnras\/} {\bf 432}, 3438 (2013).

\end{thebibliography}


\begin{thebibliography}{10}

\bibitem{dej87}
H.~{Dejonghe}, {\it \mnras\/} {\bf 224}, 13 (1987).

\bibitem{cha+02b}
P.~{Chatterjee}, L.~{Hernquist}, A.~{Loeb}, {\it \apj\/} {\bf 572}, 371 (2002).

\bibitem{gie+94}
M.~{Giersz}, D.~C. {Heggie}, {\it \mnras\/} {\bf 268}, 257 (1994).

\bibitem{mur+91}
B.~W. {Murphy}, H.~N. {Cohn}, R.~H. {Durisen}, {\it \apj\/} {\bf 370}, 60
  (1991).

\bibitem{heg+03}
D.~{Heggie}, P.~{Hut}, {\it {The Gravitational Million-Body Problem: A
  Multidisciplinary Approach to Star Cluster Dynamics}\/} (Cambridge University
  Press, 2003, 372 pp., 2003).

\bibitem{gur+07}
M.~A. {G{\"u}rkan}, C.~{Hopman}, {\it \mnras\/} {\bf 379}, 1083 (2007).

\bibitem{eil+09}
E.~{Eilon}, G.~{Kupi}, T.~{Alexander}, {\it \apj\/} {\bf 698}, 641 (2009).

\bibitem{fra+02}
J.~{Frank}, A.~{King}, D.~J. {Raine}, {\it {Accretion Power in Astrophysics}\/}
  (Cambridge University Press, 2002), {third} edn.

\bibitem{ill+00}
A.~F. {Illarionov}, A.~M. {Beloborodov}, {\it \mnras\/} {\bf 323}, 159 (2001).

\bibitem{dav+80}
R.~E. {Davies}, J.~E. {Pringle}, {\it \mnras\/} {\bf 191}, 599 (1980).

\bibitem{ruf97}
M.~{Ruffert}, {\it \aap\/} {\bf 317}, 793 (1997).

\end{thebibliography}
\end{document}